\documentclass[aps,twocolumn,showpacs]{revtex4}
\usepackage{latexsym}

\newcommand{\be}{\begin{equation}}
\newcommand{\ee}{\end{equation}}
\newcommand{\bea}{\begin{eqnarray}}
\newcommand{\eea}{\end{eqnarray}}
\newcommand{\bref}[1]{(\ref{#1})}
\newcommand{\nsp}{\hspace{-0.18in}}

\newcommand{\Lag}{\mathcal{L}}
\newcommand{\Lagb}{\mathcal{L}^{(b)}}
\newcommand{\mpl}{M_\mathrm{Pl}}
\newcommand{\ms}{M_\mathrm{S}}
\newcommand{\dz}{\partial_z}

\begin{document}

\title{Modified Large Distance Newton Potential on a Gauss-Bonnet Brane World}

\author{Stephen C. Davis}
\email{stephen.davis@epfl.ch}
\affiliation{ITP, \'Ecole Polytechnique
F\'ed\'erale de Lausanne, CH--1015 Lausanne, Switzerland}

\begin{abstract}
Gravity on a brane world with higher order curvature terms and a
conformally coupled bulk scalar field is investigated. Solutions with
non-standard large distance gravity are described. It is not necessary
to include a scalar field potential in order to obtain the
solutions. The resulting Newton potential is qualitatively similar to
that of the Dvali-Gabadadze-Porrati (DGP) model. For suitable
parameter choices the model is ghost free. Like many other brane
gravity models with modified large distance Newton potentials, the
short distance gravity is scalar-tensor. The scalar couples with
gravitational strength, and so the model is not compatible with observation.
\end{abstract}


\maketitle

\section{Introduction}

In the brane world scenario our four dimensional universe is a brane
embedded in a higher dimensional bulk space. There are many versions
of this scenario, but in this paper we will be interested in models
with a single brane in an infinite, five-dimensional bulk space. In
the Randall-Sundrum (RS) model, the warping of the bulk traps a graviton
zero mode on the brane~\cite{RS}. This allows standard four
dimensional gravity to be obtained there (at least for larger
distance scales). 
Brane models which do not give conventional gravity at larger distances
have also been proposed~\cite{DGP,GRS,CEH,bigravity,ghostcond}.
In this article we will be interested in a brane
model which does not have a
massless graviton bound state. The effective four-dimensional gravity
is obtained by a different method. Such models have a non-standard large
distance Newton potential.  There are also corrections to it at
shorter distances, which are of interest in an astrophysical
context.

One popular model with modified large distance gravity is the
Dvali-Gabadadze-Porrati (DGP) model~\cite{DGP}. It has an unwarped
bulk, so that five dimensional gravity is obtained at large
distances. Four dimensional gravity is obtained at short distances due
an induced curvature term on the brane.

We will consider a brane model which has a conformally coupled scalar field
and whose action includes the quadratic order Gauss-Bonnet curvature
term. It is just as natural to include the Gauss-Bonnet term in the
action as the usual Einstein-Hilbert term, since the resulting gravitational
field equations are as well behaved as the Einstein
equations~\cite{Lovelock}. We will also include second order kinetic
terms for the scalar field. The literature contains many papers on
linearised gravity in Gauss-Bonnet brane models~\cite{BWGB,us,me}. We
will review the relevant points in section~\ref{sec:grav}.

The boundary action associated with the Gauss-Bonnet term includes the
induced Einstein tensor~\cite{Myers}. For a warped bulk spacetime,
this will give a similar contribution to the effective
four-dimensional theory as the induced gravity term in the DGP
model. This suggests that solutions can be found with a similar
behaviour to the DGP model. We will show in section~\ref{sec:newt}
that, with the help of the scalar field, this is indeed the case. Our
solutions do not appear to have the strong interaction problems of DGP
model~\cite{DGPprob,Val}.

Closer examination of the solutions reveals that they have much in
common with quasilocalised brane gravity models~\cite{GRS,CEH}. In
these models the bulk is warped, but not sufficiently for there to be
a bound graviton zero mode. Four dimensional gravity is instead
obtained from the massive graviton states. Previous quasilocalised
models have suffered from problems with ghosts~\cite{GRSprob}. For
suitable parameter choices these can be avoided for our solutions. 

Unfortunately the bulk
gravitons in our model have extra an degree of freedom. The resulting
four dimensional theory contradicts solar system observations, as it
does not give the correct predictions for light bending. This problem
is discussed in section~\ref{sec:diss}.

In section~\ref{sec:PGB} we consider the implications of our analysis
for a model with only second order gravity terms in its action, and
show that (for the class of solutions we are considering) all
solutions are unstable.

\section{Effective Four Dimensional Gravity}
\label{sec:grav}

We will consider a brane model with the gravitational action
\bea
&&\nsp S =\frac{1}{2\kappa^2} \int \! d^5x \sqrt{-g} \,
e^{-2\Phi}\left\{ a_1 R - 4a_2 (\nabla \Phi)^2 + \Lag_2 - 2\Lambda\right\} 
\nonumber \\ && \hspace*{0.2in}
- \frac{1}{\kappa^2} \int \! d^4x \sqrt{-h}  \, 
e^{-2\Phi} \left\{ 2a_1 K + \Lagb_2 + T\right\}
\label{act}
\eea
where $\Lambda$ and $T$ are respectively the bulk cosmological constant
and brane tension, and $\Lag_2$ and $\Lagb_2$ give the second order
curvature contributions.
We are treating the brane as a boundary of the bulk space, and so have
included the Gibbons-Hawking boundary term~\cite{GibbHawk}. 
$h_{ab} = g_{ab} - n_a n_b$ is the induced metric on the brane and
$K_{ab} = h^c{}_a \nabla_{\! c} n_b$ is the extrinsic curvature.

The quadratic order part of the action is
\bea
\label{act2}
&&\!\!\!
 \Lag_2 = c_1 \left(R^2 - 4 R_{ab}R^{ab} + R^{abcd} R_{abcd}\right)
\nonumber \\  &&\hspace*{0.3 in} {}
- 16 c_2 G_{ab} \nabla^a\Phi\nabla^b\Phi
\nonumber \\  &&\hspace*{0.3 in} {}
+ 16 c_3 (\nabla \Phi)^2 \nabla^2\Phi - 16 c_4 (\nabla \Phi)^4
\eea
\bea
\label{actb2}
&&\!\!\!
\Lagb_2 = \frac{4}{3} c_1 \bigl(3 K K_{ab}K^{ab} - 2K^a{}_b K^{bc}K_{ac} 
- K^3 
\nonumber \\  &&\hspace*{0.5 in} {}
-6 G^{(4)}_{ab}K^{ab} \bigr) 
- 16 c_2 (K_{ab}- K h_{ab})D^a\Phi D^b \Phi
\nonumber \\  &&\hspace*{0.3 in} {}
+ \frac{16}{3} c_3\left((n^a \partial_a\Phi)^3 
+ 3n^a\partial_a\Phi (D\Phi)^2\right) \ .
\eea
Again, appropriate boundary terms have been included~\cite{me,Myers,BCGB}.
The full bulk field equations and brane junction conditions obtained
from the action~\bref{act} are given in ref.~\cite{me}. Note that
$G_{ab}$ denotes the Einstein tensor, and not the induced metric.

We will be interested in perturbations of a RS-like background, with
$ds^2= e^{2A(z)}(dx^2_4 + dz^2)$, $\Phi'/A' =u = \mathrm{constant}$ and
$A(z) = -\ln(1+|z|/\ell)$. We take $u$ to be constant for
simplicity. Furthermore, the form of the resulting solutions is similar to the
RS and DGP brane worlds, allowing easy comparison of the different models.
The brane is at $z=0$ and the
bulk is taken to be $Z_2$ symmetric. We take $u$ and $\ell$ to be
positive, as we are interested in RS-like solutions without bulk
singularities. The relations between the
different parameters are listed in the appendix. 

We consider the perturbed metric
\bea
&& ds^2 = e^{2A}\left[
(\eta_{\mu \nu} + \gamma_{\mu \nu}) dx^\mu dx^\nu + dz^2 \right] 
\nonumber \\ && \hspace*{0.4in}
- \ell e^A dz (dx^\mu \partial_\mu + dz\dz) 
\left( N_1\psi + \frac{2}{u}\varphi \right)
\label{pertmet}
\eea
and take $\Phi = \phi_0 + u A + \varphi$. The metric perturbation is expanded as
\be
\gamma_{\mu \nu} = 
\bar \gamma_{\mu \nu} + 2(\zeta+\psi) \eta_{\mu \nu} 
-2N_2 \frac{\partial_\mu \partial_\nu}{\Box_4} \psi
\label{gdef}
\ee
where $\partial^\mu \bar \gamma_{\mu \nu} = 0$ and 
$\eta^{\mu \nu}\bar \gamma_{\mu \nu} = 0$. The fields $\psi$ and
$\zeta$ are linear combinations of $\varphi$ and 
$\gamma= \eta^{\mu \nu} \gamma_{\mu \nu}$. We work in a gauge in which
the brane does not bend, even when matter is added to it (see e.g.\
ref~\cite{me} for more details). This is
achieved by suitable choices of $\zeta$, $\psi$ and the coefficients
$N_1$, $N_2$ (see appendix).

The linearised bulk field equations give 
\be
\mu_\gamma \left(\dz^2 - (3 -2u) \ell^{-1} e^A \dz 
+ f_\gamma^2 \Box_4\right)\bar \gamma_{\mu \nu} = 0
\label{Bg}
\ee
where
\be
\mu_\gamma = a_1 -4\ell^{-2} (c_1 (1-4u) + 2c_2u^2)
\ee
\be
f^2_\gamma = 1 + 8u\frac{c_1(1+2u)+2c_2u}{\mu_\gamma\ell^2} \ .
\ee
If either of $\mu_\gamma$ or $f^2_\gamma$ are negative, the model
will have graviton ghosts. This requirement restricts allowable ranges
of the parameters $a_i$ and $c_i$ in the action. The graviton wave
equation~\bref{Bg} is solved (for spacelike momenta $p$) by
\be
\bar \gamma_{\mu \nu} \propto 
e^{-(2-u)A} K_{2-u}\left(f_\gamma p \ell e^{-A}\right) \ .
\label{gsol}
\ee
where $K_{2-u}$ is the order $(2-u)$ modified Bessel function of the
second kind. The effective gravitational law on the brane is given by
the junction conditions. We find
\be
2\mu_\gamma \dz \bar \gamma_{\mu \nu}+m_\gamma^2  \Box \bar \gamma_{\mu \nu}
= -2\kappa_*^2  \left\{S_{\mu\nu} 
- \frac{1}{3}\left(\eta_{\mu\nu} 
- \frac{\partial_\mu \partial_\nu}{\Box_4}\right)S \right\}
\label{bcg}
\ee
where $S_{\mu\nu}$ is the perturbation of the energy momentum tensor
on the brane and $\kappa_*= \kappa e^{\phi_0}$. The parameter
$m_\gamma^2 = 8 c_1(1-2u) /\ell$ must be positive, or the theory will
have tachyonic graviton modes~\cite{me,us}.

The behaviour of the scalar field $\psi$ is qualitatively similar. Its
bulk field equation is the same as eq.~\bref{Bg}, but with
$\mu_\gamma$ and $f^2_\gamma$ replaced by $\mu_\psi$ and
$f^2_\psi$. Its junction condition is
\be
2\mu_\psi \dz \psi + m_\psi^2 \Box_4 \psi = -\kappa_*^2 S \ .
\label{bcp}
\ee
Again ghosts and tachyons are present for some parameter ranges.

The remaining degree of freedom, $\zeta$, is pure gauge in the
bulk, but its behaviour on the brane is fixed by the junction conditions
\be
m_\zeta^2 \Box_4 \zeta = - \kappa_*^2 S \ .
\ee
If we require that $\zeta$ is not a ghost, $m_\zeta^2$ must be positive.

Requiring that the model is completely free from ghosts and tachyons
will rule out large regions of parameter space, although solutions 
which do not suffer from either of these types of
instabilities do exist. An example is given at the end of the next section.

\section{Modified Newton Potential}
\label{sec:newt}

By substituting the solution~\bref{gsol} into the junction conditions
we can obtain the linearised effective four dimensional gravity. If we
then consider a perturbation corresponding to a point mass at $r=0$ on
the brane, we can also determine the effective Newton potential $V_N$. The
motion of a test particle confined to the brane is given by
$d^2x^i/dt^2 \approx -{}^{(4)}\Gamma^i_{00}$, and from this we obtain
$V_N = (1/2)\gamma_{00}$.

If $0 \leq u <1$ then for small $p$ (which corresponds to large
distance scales) a series expansion of eq.~\bref{gsol} gives
\be
\dz \bar \gamma_{\mu\nu} \approx 
-\frac{p^2 f^2_\gamma\ell}{2(1-u)}\bar \gamma_{\mu\nu} \ .
\label{bigr1} 
\ee
Thus the junction condition~\bref{bcg} implies $\bar \gamma_{\mu \nu}
\propto \{S_{\mu\nu}- \cdots\}/p^2$. Fourier transforming this when
$S_{\mu\nu}$ corresponds to a point mass [so $S_{00} \propto
\delta(p^0)$], we find it gives a standard $1/r$ contribution to
Newton's law. This is qualitatively identical to what happens in the
RS model (which can be obtained from the above case by taking
$u=0$). Alternatively, if we look at the graviton spectrum for this
model, we will see that it has a zero mode which is localised on the
brane. This mode gives the dominant contribution to the large distance
gravity of this type of solution.

If $1<u<2$ then the expression~\bref{bigr1} is no longer valid, and instead
we have
\be
\dz \bar \gamma_{\mu\nu} \approx -\frac{2\Gamma(u-1)}{\ell\Gamma(2-u)}
\left(\frac{p\ell f_\gamma}{2}\right)^{4-2u} \bar \gamma_{\mu\nu} 
\label{bigr2} 
\ee
at large distances. Substituting eq.~\bref{bigr2} into the junction
condition~\bref{bcg}, we again obtain $\bar \gamma_{\mu\nu}$ in terms of the
energy momentum perturbation $S_{\mu \nu}$. We now find that it gives a
non-standard $1/r^{2u-1}$ contribution to the large distance Newton
potential. This is the same as would be obtained for a
$2(u+1)$-dimensional theory. A special case of this is $u=3/2$, for which 
$\dz \bar \gamma_{\mu\nu} = -f_\gamma p \bar \gamma_{\mu\nu}$ at all
scales. The large distance contribution to the Newton potential is
then $1/r^2$, which is the same as is obtained in the DGP model.
The scalar mode, $\psi$, gives similar non-standard contributions to the
large distance Newton's law.

We now have only massive graviton bound states and no
localised zero mode. However we can still obtain four-dimensional
gravity at short distances, since in this case the $m_\gamma^2 \Box_4
\bar \gamma_{\mu \nu}$ term dominates the junction
condition~\bref{bcg}, and (to leading order) a standard $1/r$ Newton
potential is obtained.

Putting all the junction conditions together, we obtain the following
expression for the induced metric perturbation
\bea
&& \nsp \gamma_{\mu \nu}(p) = 2 \kappa_*^2\Biggl(
 D_\gamma(p) \left\{S_{\mu\nu} 
- \frac{1}{3}\left(\eta_{\mu\nu} 
- \frac{p_\mu p_\nu}{p^2}\right)S \right\}
\nonumber \\ && \hspace{0.2in} {}
+ D_\psi(p) \left\{ \eta_{\mu\nu} 
- N_2 \frac{p_\mu p_\nu}{p^2}\right\}S
+\eta_{\mu\nu} \frac{S}{m_\zeta^2 p^2}
\Biggr)
\label{prop}
\eea
where
\be
D_i(p) = \left\{m_i^2 p^2 + 2\mu_i f_i p 
\frac{K_{u-1}(f_i \ell p)}{K_{2-u}(f_i \ell p)} \right\}^{-1} \ .
\ee
The three contributions to eq.~\bref{prop} correspond respectively to
the bulk graviton modes, the bulk scalar, and the `brane-bending' mode. 
The Newton potential, $V_N$, and the effective four-dimensional graviton
propagator can be extracted from the above expression~\bref{prop}.

Defining $r_i = m^2_i /(2\mu_i f_i)$, we find that for large momentum
$D_i^{-1}(p) = m_i^2[p^2 + p/r_i+O(1)]$. Using this asymptotic behaviour
of $D_i(p)$, and taking the perturbation $S_{\mu\nu}$ to be a point mass,
we can determine $\gamma_{00}$. By fourier transforming this we can obtain
the approximate short distance Newton potential
\bea
&&\nsp V_N = -\frac{\kappa_*^2}{4\pi r}
\left(\frac{2}{3m^2_\gamma}+\frac{1}{m^2_\psi}+\frac{1}{m^2_\zeta}\right)
\nonumber \\ && {}
-\frac{\kappa_*^2}{2\pi^2}\left\{
\frac{2}{3r_\gamma}\ln \left(\frac{r}{r_\gamma}\right)
+\frac{1}{r_\psi}\ln \left(\frac{r}{r_\psi}\right) \right\} + O(1) \ .
\hspace*{0.25in}
\eea
which will be valid when $r \ll r_\gamma, r_\psi, \ell f_\gamma, \ell f_\psi$.
The logarithmic corrections are similar to those that appear in the DGP
model. In fact, they will be present even for solutions with localised
($0\leq u<1$) gravity. Similar corrections also occur in warped
space versions of the DGP model~\cite{WDGPgrav}.

In a similar way we can obtain the large distance effective gravity by using
a power series expansion of $D_i(p)$ in the expression for
$\gamma$~\bref{prop}. If $1<u<2$, then for large distances
($r \gg  r_\gamma, r_\psi, \ell f_\gamma, \ell f_\psi$)
\bea
&& \nsp V_N = -\frac{\kappa_*^2}{4\pi m^2_\zeta r}
- \frac{\kappa_*^2\Gamma(u-1/2)}{4\pi^{3/2}\Gamma(u-1)}
\nonumber \\ && \hspace*{0.4in} {}
\times \left(
\frac{2}{3\mu_\gamma f^{4-2u}_\gamma} + \frac{1}{\mu_\psi f^{4-2u}_\psi}
\right) \frac{\ell^{2u-3}}{r^{2u-1}}
\label{VNbig}
\eea
to leading order. Note that the power series expansion of $K_{u-1}$
used to derive eq.~\bref{VNbig} is only valid for $u>1$, and so
putting $u=0$ in the above expression does not give the correct Newton
potential for the RS model.

If $m_\zeta^2$ is large, the resulting large and
short distance gravity resembles the DGP model, especially for the
special case of $u=3/2$. We then have $1/r^2$ contributions to the
large distance Newton's law, as would normally occur in
five-dimensional gravity. However, even if $m_\zeta^2$ is very large, the first
term in the above expression~\bref{VNbig} will eventually dominate. We
will therefore have a $1/r$ Newton potential at very large distances, in
contrast to the DGP model. At intermediate scales, we will have a
combination of logarithmic and $1/r^{2u-1}$ corrections.

One problem with previous quasilocalised brane gravity models is that
they have ghosts~\cite{GRSprob}. For a wide range of parameter choices, our
solutions have the same problem, although there are exceptions.  For
example suppose $u$ is fairly close to 1 ($1< u \lesssim 1.07$) and
the parameters in the action~\bref{act} are  $c_2= (7-u) c_1/4$, 
$c_3 = (13-73u+126u^2-18u^3)c_1/(16u^3)$, 
$c_4= (13-64u+60u^2+41u^3-14u^4)c_1/(8u^4)$, $a_1= (u-1)(8u-7)c_1/\ell^2$
and $a_2=0$. If $c_1$ is negative, then $m^2_i, \mu_i >0$ and the
solution is free from ghosts and tachyons. To leading order in $u-1$
we find $\mu_\gamma\ell \sim m_\gamma^2 \sim m_\zeta^2 \sim
-(u-1)c_1/\ell$, $f_\psi^2 \sim(1-u)^2$, $f_\gamma^2 \sim 1$,
$m_\psi^2 \sim -c_1/\ell$ and  $\mu_\psi \sim -c_1/\ell^2/(u-1)$.
Hence we have shown (by construction) that ghost free solutions with a
non-standard large distance Newton's law can be found. The above solution also
shows that a modified Newton's law which is very close to the usual
$1/r$ four dimensional one can be found.

Although the DGP model does not have ghosts, it does have strong
interactions at short distances, raising doubts about the model's
predictivity~\cite{DGPprob,Val}. It was
shown in ref.~\cite{DGPprob} that the strong interaction scale comes from
the small kinetic term for the brane bending mode. In our model this
is $m_\zeta^2$, which is not generally small, suggesting that there
need not be a corresponding strong interaction problem for our model.

\section{Relativistic Effects}
\label{sec:diss}

While our model does give the correct Newton potential at short
distances, this is not sufficient for it to be compatible with
gravity measurements of the solar system.
At short distances ($1/p \ll \ell f_{\gamma,\psi}, r_{\gamma,\psi}$)
the leading order behaviour of the perturbation~\bref{prop} is
\be
\gamma_{\mu \nu}(p) \approx
\frac{2}{\mpl^2 p^2} \left\{S_{\mu\nu} - \frac{1}{2}\eta_{\mu\nu} S \right\}
+ \frac{1}{\ms^2 p^2}\ \eta_{\mu\nu} S \ .
\label{propBD}
\ee
We have omitted the $p_\mu p_\nu/p^2$ dependent terms for simplicity.
This describes scalar-tensor gravity, with Planck mass $\mpl = 
m_\gamma/\kappa_*$ and scalar mass $\ms$ given by
\be
\frac{1}{\ms^2} = 2\kappa_*^2 \left(\frac{1}{6m^2_\gamma} 
+ \frac{1}{m_\psi^2} + \frac{1}{m_\zeta^2}\right) \ .
\label{Ms}
\ee
To avoid conflict with solar system gravity tests, we need
$\ms \gg \mpl$~\cite{vDVZ,Damour}. This is not possible if
$m_\zeta^2$ and $m_\psi^2$ are both positive.

The problem arises because the contribution from the $\bar
\gamma_{\mu\nu}$ modes [the first term of eq.~\bref{prop}] has the
wrong tensor structure. The presence of the factor $1/3$, as opposed
to $1/2$, indicates that there is an extra degree of freedom, the
graviscalar, which couples to the energy-momentum tensor. A similar
situation occurs in massive four-dimensional gravity.  The effects of
this extra graviscalar are not compatible with
observations~\cite{vDVZ}. Our model also has the additional $\psi$ and
$\zeta$ scalar modes. These will make things even worse, although
their effects could be suppressed by taking $m_\zeta^2, m_\psi^2 \gg
m_\gamma^2$. A similar problem occurred with quasilocalised gravity
models~\cite{GRSprob}. This was originally thought to be a problem
with the DGP model, although the situation is actually more
complicated due to the model's strong interactions. In general, it
seems that our model does not have this feature, suggesting its
predictions are more reliable. However since these predictions appear
to be incompatible with observations, this is not very useful.

To obtain an effective four dimensional theory with correct
relativistic behaviour, the contribution of the extra graviscalar must
be cancelled, either with the brane-bending field, $\zeta$, or the bulk
scalar, $\psi$.  This is not possible unless we have either $m_\zeta^2$
(or $m_\psi^2$) is negative, as was the case for quasilocalised gravity
models. Looking at the linearised action for these scalar fields
\bea
&& -\frac{1}{2\kappa_*^2}\int d^5x \, e^{(3-2u)A} \mu_\psi 
\left\{ (\dz\psi)^2 +f_\psi^2 (\partial_\nu \psi)^2\right\} 
\nonumber \\ && \ \ \ {}
- \frac{1}{2\kappa_*^2}\int d^4x \left\{ m_\psi^2 (\partial_\nu \psi)^2 
+ m_\zeta^2 (\partial_\nu \zeta)^2 \right\}
\eea
we see that if $m_\zeta^2<0$ the theory will have a ghost. If
$m_\psi^2<0$ then either $\psi$ is a ghost (if $\mu_\psi <0$), or it
has a tachyon mode [since eq.~\bref{bcp} has solutions with spacelike
momenta even in the vacuum].

In the RS model $m_\zeta^2$ is negative, but this is not a problem. This
model has two massless graviton zero modes, and an unphysical
graviscalar zero mode. The brane-bending ghost zero mode cancels the
graviscalar (a similar idea is used in the quantisation of QED).
This does not work for quasilocalised gravity models since there
are no massless graviscalar or scalar states for the ghost to cancel
with~\cite{GRSprob}. Furthermore, the resulting Newton potential is
repulsive at large $r$.

For our model it is possible to cancel the
contribution of the massive graviscalar modes by making $\psi$ a
ghost. In general this is likely to be a physical ghost, in contrast
to $\zeta$ in the RS model. However if $\psi$ and 
$\bar \gamma_{\mu\nu}$ have exactly the same spectrum, this may not be the
case. For this we need $f_\psi =f_\gamma$ and 
$\mu_\psi/m^2_\psi= \mu_\gamma/m^2_\gamma$. To get $\ms \gg \mpl$ we
also need $m^2_\psi \approx -m^2_\gamma/6$ and $m_\zeta^2 \gg m^2_\gamma$. 
The resulting effective four dimensional gravity would be compatible
with observations, at least to linear order. To know if $\psi$
is a physical ghost or not, we would have to go to higher order in
perturbation theory, and calculate interaction terms.

In fact, having $m_\psi^2$ negative may not be as bad
as it seems. Normally such a ghost field would imply that the energy
of the theory is unbounded from below. However, the higher order terms
in the action mean that terms such as $(\partial \psi)^4$ will also
be present. If the effective action contains a term such as
$-C_1(\partial_\mu \psi \partial_\nu \psi-C_2\eta_{\mu\nu})^2$, with
$C_1,C_2 >0$, then $m_\psi^2$ is negative, but the theory's energy is
still bounded from below.

Throughout this article we have assumed that a linearised analysis is
sufficient to determine the effective gravity in our model. If any of
$m_i$ are zero or very small, this will not be the case and the
expression for $\ms$~\bref{Ms} will not be valid. It is possible that by
going to quadratic order we will obtain brane gravity that more
closely resembles general relativity at shorter distances. A similar
situation arises in the model considered in ref.~\cite{padilla}, where
conventional gravity was obtained at intermediate length scales when
the analysis was extended beyond linear order.

Even if there is no viable solution to the problem of the extra
degrees of freedom in our modified gravity model, the above analysis can
still be used to impose constraints on the parameters of the underlying
theory. For the action~\bref{act} there are still solutions with $u<1$
and RS-style localised gravity. This is in contrast to the
corresponding first order gravity theories, which have no solutions
with localised gravity.

\section{Pure second order gravity}
\label{sec:PGB}

If we remove either the first or second order terms from the action,
then the algebraic expressions for the various coefficients in the
previous sections simplify considerably. If $a_i=0$ then
\be
f^2_\gamma = -1 + \frac{\ell m^2_\gamma}{\mu_\gamma}(u-1) \ . 
\ee
To avoid graviton tachyons and ghosts we need $\mu_\gamma$, 
$m^2_\gamma=8c_1(1-2u)/\ell$ and $f_\gamma^2$ to be positive. This is
not possible unless $u>1$ and $c_1, c_2  < 0$. Thus we cannot get
stable solutions with localised gravity unless both first and second
order terms are included in the action.

Furthermore
\be
\mu_\psi f_\psi^2 = -6 \ell \mu_\gamma  T
\frac{T\ell f_\gamma^2+6(u-1)\mu_\gamma}{(\ell T- 6\mu_\gamma)^2} \ .
\ee
To have a ghost free theory we require $m_\zeta^2 = \ell^2 T$,
$f_\gamma^2$, $\mu_\gamma$ and the above expression to all be
positive. Since $u>1$, this is not possible for any choice of
parameters and so the theory always has ghosts.

As we discussed in the previous section, if $\psi$ is a ghost, it is
conceivable that it could cancel with the graviscalar. For this to
work we would need $f_\psi=f_\gamma$ and
$\mu_\psi/m^2_\psi=\mu_\gamma/m^2_\gamma$. If $\mu_\gamma>0$ this
occurs when $c_3 = (2c_2/c_1-1)c_2$ and $c_4=c_3c_2/c_1$. However this
is not acceptable since in this case $m_\zeta^2=2\ell\mu_\gamma (3-2u)
<0$, and so we have a massless, physical ghost.

Hence we can rule out all pure second order gravity solutions, since
they all have ghosts or tachyons. Pure first order gravity (with a
scalar field) has similar problems.

\section{Summary}

We have investigated the effective four-dimensional gravity of a brane
world scenario with a conformally coupled scalar field and quadratic
order curvature terms. As well as localised gravity solutions which
give a conventional large distance Newton potential, there also exist
solutions with modified large distance gravity.

These solutions have features of both the DGP and quasilocalised
gravity scenarios. The bulk space is warped, but we can still obtain a
DGP-like Newton's law, and do not appear to have the strong interaction
problems of that model. The theory's higher curvature terms allow a
short distance Newton's law with $1/r$ behaviour to be obtained. It
has logarithmic corrections like those in the DGP model.  The large
distance Newton's law includes $1/r^{2u-1}$ terms, with $1<u<2$. For
$u=3/2$ this resembles the large distance behaviour of the DGP model.

Examination of the tensor structure of the graviton
propagator reveals that at short distances the effective gravity of the
solution is scalar-tensor, and that the scalar and gravitational
couplings are of similar strength. This is incompatible with solar system
gravity experiments. It may be possible to fix the problem by making the bulk
scalar field a ghost, although the resulting theory is unlikely to be
consistent.

For the special case of only second order curvature terms (and
corresponding scalar kinetic terms), all
solutions have either physical ghosts or tachyons. We see that when a
scalar field is included in the model, solutions with well behaved
effective gravity can only be found if the bulk action includes both
first and second order curvature terms.

\section*{Acknowledgements}

I wish to thank Valery Rubakov, Mikhail Shaposhnikov, Katarzyna
Zuleta Estrugo for useful comments. I am grateful to the Swiss
Science Foundation for financial support.

\appendix
\section{Parameters}

The bulk cosmological constant and brane tension are respectively
$\Lambda=\Lambda^{(1)}+\Lambda^{(2)}$ and $T=T^{(1)}+T^{(2)}$ where
\be
\Lambda^{(1)} = \frac{2}{\ell^2}(a_1(4u-3) + a_2 u^2)
\ee
\be
\Lambda^{(2)} = \frac{4}{\ell^4}(3c_1(1-8u) +36c_2u^2-4c_3(4+u)u^3+6c_4u^4)
\ee
\be
T^{(1)} = \frac{2}{\ell}(3-2u)a_1
\ee
\be
T^{(2)} = -\frac{8}{3\ell^3}(3c_1(1-6u) + 18c_2 u^2 - 4c_3 u^3) \ .
\ee
We will find it useful to define $\Lambda_+ = 2\Lambda^{(1)}+4\Lambda^{(2)}$ 
and $T_\pm = \pm T^{(1)} + 3T^{(2)}$.

The parameter $u=\Phi'/A'$ satisfies the equation
\bea
&&\nsp a_1 + 2 (a_1+a_2) u - 4\ell^{-2} [
3c_1 + 6(c_1-2c_2)u 
\nonumber \\ && \hspace{0.3in} {}
+ 2(4c_3-3c_2)u^2 + 4(c_3-c_4)u^3]  = 0 \ .
\label{ic}
\eea
In the expressions for the metric
perturbation~(\ref{pertmet},\ref{gdef}), the coefficients are
\be
N_1 = 12
\frac{\mu_\gamma (T_+\ell + 4a_1[2-3u])}{T_+\ell (T_+\ell+8a_1u-6\mu_\gamma)}
\ee
\be
N_2 = \left(\frac{\Lambda_+}{T_+} + \frac{T_+}{3\mu_\gamma}\right)N_1 \ell
\ee
and the scalar fields are
\be
\psi = \frac{T_+}{2N_1 \ell \Lambda_+ u} (8\varphi - u \gamma)
\ee
\be
\zeta = \frac{\varphi}{u} 
+ \frac{6(\mu_\gamma-2a_1)}{T_+\ell+8a_1u-6\mu_\gamma} \psi \ .
\ee
The coefficients in the wave equations for these fields are
\be
\mu_\psi = \frac{N_1^2\ell^2}{4} 
\left(\Lambda_+ +\frac{T_+^2}{3\mu_\gamma}\right)
\ee
\be
f_\psi^2 = -\frac{N_1^2 T_+\ell}{4 \mu_\psi}
\left(\frac{\ell T_+}{6\mu_\gamma} f_\gamma^2 + 1 - u\right)
\ee
\be
m_\psi^2 = -\frac{T_+ N_1^2 \ell^2}{4}
\left( 1 + \frac{T_+ m_\gamma^2}{6\mu_\gamma^2}
+ \frac{T_+(\mu_\gamma-2a_1)^2}{T_- \mu_\gamma^2}\right)
\ee
\be
m_\zeta^2  = \ell^2 T_- \ .
\ee

%some scalar gravity hep-th/0307206

\end{document}